# Alternative security architecture for IP Telephony based on digital watermarking

Wojciech Mazurczyk[1], Zbigniew Kotulski[1,2]

[1]Warsaw University of Technology, Faculty of Electronics and Information Technology,
Institute of Telecommunications
`{wmazurczyk, zkotulsk}@tele.pw.edu.pl`
[2]Polish Academy of Sciences, Institute of Fundamental Technological Research
`zkotulsk@ippt.gov.pl`

**Abstract.** Problems with securing IP Telephony systems, insufficient standardization and lack of security mechanisms emerged the need for new approaches and solutions. In this paper a new, alternative security architecture for voice-systems is presented. It is based on *digital watermarking*: a new, flexible and powerful technology that is increasingly gaining more and more attention. Besides known applications e.g. to solve copyright protection problems, we propose to use *digital watermarking* to secure not only transmitted audio but also signaling protocol that IP Telephony is based on.

## 1  Motivation: IP Telephony security problems

Most people think that main problems with IP Telephony are connected with providing Quality of Service (QoS) parameters. This is only part true. Security considerations are very important too. This does not only mean security of conversation between two communicating parties but also security of signaling messages used to make this call possible. Both those problems can be responsible for increasing latency. If it's level rises to a certain value, it can be the most degenerating constrain for IP Telephony quality of call. So, now we are always facing necessity of trade off between providing security and low latency.

IP Telephony security problems can be divided into three groups:
- Security of a signaling protocol on which such a system is based on,
- Security of the data stream (audio); for IP Telephony there are often RTP packets,
- Security connected with IP environment and networks based on this network protocol.

We would like to emphasize that it is not enough to protect the media stream that is exchanged between calling parties. Such a thinking leads to situation in which two entities are able to talk to each other in a secure manner but they are unable to initiate a call because of an attacker actions on the signaling protocol. In [8], [9] and [10] different security issues and gaps for VoIP are presented.

Signaling protocols like SIP and H.323 are often called a "heart" of Voice over Internet Protocol because they are responsible for every aspect of the future call and even for network architecture. Without proper security of the IP Telephony signaling protocols we cannot have a secure voice communication system. Classical architecture for existing IP Telephony signaling protocols defines a set of security mechanisms which we can be generally divided into two groups: **internal** (mechanisms defined especially for this signaling protocol) and **external** (mechanisms like IPSec or TLS). Combination of these mechanisms is supposed to provide us security. But this is not all so simple and we try to discuss this situation below.

To sum up our motivation section we try to find another way to handle IP Telephony security based on the following facts:
- There are drawbacks of SRTP protocol which are discussed in [3]; SRTP is the most popular mechanism to provide authentication and integrity for data stream,
- There is no "one size fits all" solution for protecting audio and signaling messages [8] for IP Telephony systems. To properly protect such systems we must consider a lot of aspects and posses a certain level of IT knowledge,
- Current security mechanisms for signaling protocols do not quite "fit" for securing real-time services like VoIP [8],
- Current security mechanisms can cause a situation in which certain network equipment on the communication path that do not support those mechanisms can break down the connection,
- There are no standardized security solutions for interworking between different signaling protocols, e.g. SIP and H.323,
- Security mechanisms for signaling protocols are often not implemented by vendors because they are increasing the cost of the network equipment (more processing time is needed),
- Standards (e.g. for SIP or H.323) define security mechanisms but often they are only optional to implement and vendors take full advantage of it.

That's why in this paper we are proposing a different approach to the security, based on digital watermarking. It is suitable to protect audio content and (or) signaling messages.

## 2   Introduction: Digital Watermarking

Digital watermarking is a multidisciplinary field deployed widely in last decade. It covers a large field of different aspect (from cryptography to signal processing) and is suitable for marking digital content. For our purposes the most interesting feature of watermarking is the ability to create a hidden channel in which we can exchange security information. Moreover this embed information is imperceptible to third party. In this context, watermarking scheme consist of the following stages, which are shown in Fig. 1.

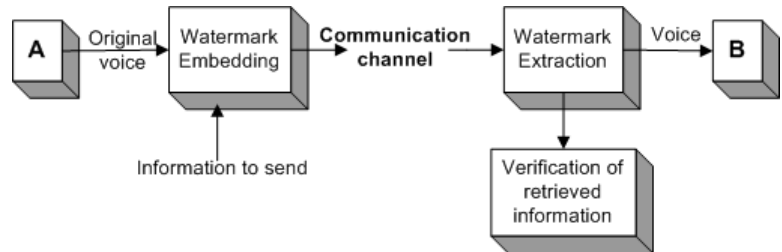

Figure 1 General watermarking scheme with information embedding

Watermark that will be used in our security architecture must posses few important properties like robustness, security, transparency, complexity, capacity and verification. All those parameters are well defined in [1] and [2]. Their optimization for real-time audio system is crucial although they are often mutually competitive so there is always a compromise necessary. Not every watermarking technique is applicable for our solution. IP Telephony is a real-time service that is why we need watermarking schemes that really work for real-time conversations. Such algorithms are described in [2], [3] and [5].

Our security architecture is watermarking scheme independent: whatever algorithm for real-time communication is used we assume that an output watermark that it created has the best properties that are allowed in this certain environment. In this way we gain flexibility of our solution and we can support future ideas and digital watermarking schemes.

## 3  New security architecture proposal

The most important security services for IP Telephony are: **authentication**, **integrity** and **confidentiality**. First two can be provided with use of watermarking techniques. The third should be guaranteed in different manner e.g. with a use of security mechanisms from classical architecture.

Making a call in IP Telephony systems consist of two phases: **signaling phase** in which certain signaling messages are exchanged between parties and **conversation phase**. Each phase has it's own, disjunctive set of security mechanisms. Nowadays, e.g. in SIP and H.323 we have different security mechanisms designated for securing data stream (audio) and different that cover signaling protocol security. Moreover the security architecture for SIP and H.323 is also different, which means that they use almost disjunctive set of security mechanisms. We are witnessing situation in which signaling and audio is secured with different mechanisms, so they are consuming a lot of processing time and can be devastating for conversation quality.

What we are proposing here is to move providing security of signaling messages from first phase to second (conversation phase). That's why we named this method a *post factum* method. Checking security of the signaling protocol is made after the signaling phase is finished. It has some disadvantages – most serious is that potential attack on signaling protocol is detected after some time during a conversation. But on the other hand this approach has certain advantages:
- it provides one a unified solution for audio and signaling protocol security,
- it is signaling protocol independent solution,
- it prevents doubling latency and excessive consuming of processing time,
- it reduces complexity (and cost!) of network equipment on the communication path,
- it can help to solve security mechanisms interworking problems between IP Telephony systems based on different signaling protocols,
- it can also help to protect audio and signaling protocol when interworking between VoIP systems and PSTN,
- unlike the existing security mechanisms, it provides **real** end-to-end security.

From the arguments above we conclude that this solution can be beneficial for IP Telephony security and can improve it's security architecture.

### 3.1 Modification of existing digital watermarking scheme

To make our proposition works, we need to modify watermarking system shown in Fig. 1. We add a new functional block called Preprocessing Stage (PPS). It will be responsible for preparing data before watermark embedding stage. Modified scheme with this new block is shown on the figure below:

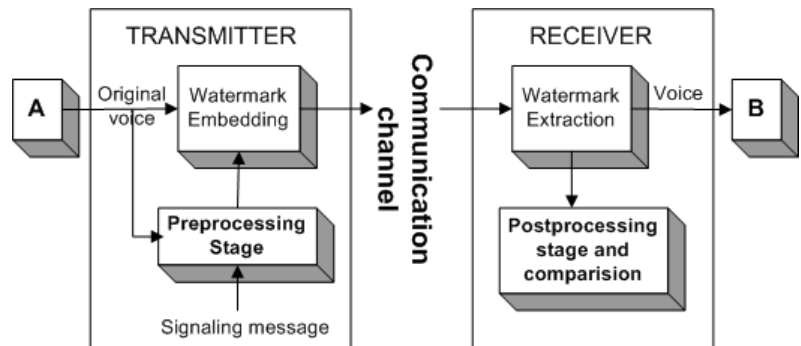
Figure 2 Modified watermarking scheme

In this system we provide as input to PPS block a signaling message and a sample of original voice. After the digital watermark is embedded and sent, in the receiver the information is retrieved and verified.

It is also important how much information we can embed into the original voice data. It defines a capacity parameter of digital watermark. This influences the speed of authentication and integrity process. In [2] different watermarking schemes developed at the Fraunhofer IPSI and the Fraunhofer IIS were considered. In our solution we need a very robust watermark to be sure that we can compare and verify sent information successfully. The result of the test in [2] shows that simultaneous large capacity and robustness depends on the scale of the codec compression. When the compression rate is high (1:53) the watermark is robust only when we embed about 1 bit/s . With lower compression rate we can obtain about 30 bit/s and the highest data rate was 48 bit/s with good robust, transparent and complexity parameters.

A speed of authentication and integrity process in our solution is expected to be high but it is not crucial. With low compression rates we propose to add pre-conversation stage. In this stage there will be few second of RTP packets exchange without conversation. It will delay the setup of the call but then during the conversation the time of verification will be shorter.

Now we will present in greater details how the Preprocessing Stage block is built. It consists of functional blocks shown in Fig. 3.

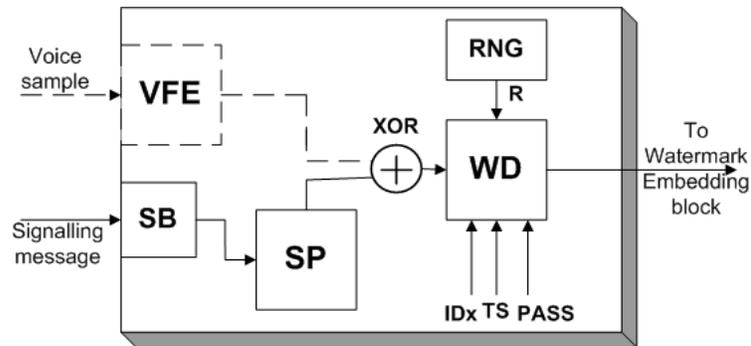

Figure 3 Preprocessing stage block

As we see in the picture this solution can be suitable for voice and(or) signaling messages protection. This gives us flexibility – we can choose if we want to secure signaling protocol, voice or both.

The blocks presented in the picture have the following functions:

**SB** (Signaling Message Buffer): stores signaling messages from the first phase of the call.

**SP** (Signaling Processing): in this block hash function is performed on each signaling message.

**RNG** (Randomizer): we use it to provide unique data for watermark even if the rest of information provided will be again the same (e.g. all signaling messages were verified so we use last sent).

**WD** (Watermarking Data): in this block input data are concatenated with obligatory R parameter and optional parameters: IDx (unique, global identifier of one side of the connection), TS (time stamp), PASS (shared password known to both sides of conversation).

This block is optional. It must be used if we want to provide integrity and authentication of both voice and signaling protocol:

**VFE** (Voice Feature Extractor): provides characteristic features of the original voice that we want to protect.

### 3.2 How the new security architecture works

The general idea of the new proposal is to compare received tokens with appropriate one, locally calculated. Fig. 4 shows how the algorithm works (if A wants to talk to B) for securing signaling messages only. It is only to simplify presentation of the solution. Securing voice is analogous.

We assume that in signaling phase some signaling messages were exchanged and now in the second phase they will be verified. $SM_S$ or $SM_R$ means Signaling Message Sent/Received. **H** means hash function and **W** is the embedding digital watermark into audio.

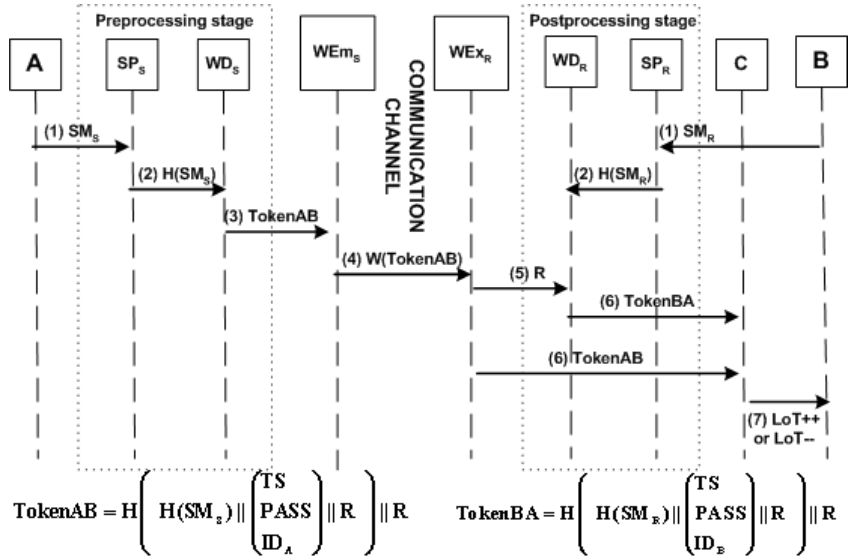

$$TokenAB = H\left(H(SM_S) \| \begin{pmatrix} TS \\ PASS \\ ID_A \end{pmatrix} \| R\right) \| R \qquad TokenBA = H\left(H(SM_R) \| \begin{pmatrix} TS \\ PASS \\ ID_B \end{pmatrix} \| R\right) \| R$$

Figure 4 Algorithm for securing signaling messages

If we want to take a full advantage from this solution and provide security of the voice the tokens changes adequately:

$$TokenAB = H\left(H(SM_S) \| \begin{pmatrix} TS \\ PASS \\ ID_A \end{pmatrix} \| R \| VF_S\right) \| R$$

$$TokenBA = H\left(H(SM_R) \| \begin{pmatrix} TS \\ PASS \\ ID_B \end{pmatrix} \| R \| VF_R\right) \| R$$

Where $VF_S$ and $VF_R$ means Voice Feature Sent/Received.

We see that securing process depends on exchanging security tokens that are calculated as shown in the Fig. 4. Then in the receiver tokens are compared with locally calculated and if they are equal a LoT (Level of Trust) parameter is increased. In any other situation LoT value decreases. For called party B the algorithm of handling the LoT parameter works as described below in a pseudo-code:

```
START
/* CL - Critical Level, LoT - Level of Trust, T - timer */
CL = a; LoT_A = x; T_A = 0; /* Initiating values */
StartTimer(T_A);
FOR (i = 0; i++; i< End of Transmission)
/* i - Time slot */
{
IF (TokenAB_i = TokenBA_i) THEN
   {
   LoT_A ++;
   ResetTimer(T_A);
   }
   ELSE (LoT_A --);
IF (LoT_A <= CL) OR (T_A > k) THEN STOP;     (1)
IF (LoT_A = a*x) THEN LoT = x;               (2)
}
```

We can see that breaking of the call will happen if the value of LoT parameter is equal or below given threshold (CL value) or if the timer $T_A$ expires (1). If the communication continues and every signaling message that was sent is verified embedding of the digital watermark do not stop. It is a continuous process: to calculate information to send after all signaling messages were verified we take last signaling message. The LoT value changes during the conversation time and if every signaling message was successfully verified the LoT value is rising. To prevent it's increasing to infinity we lower it after it reaches a value of critical level multiply by start value of LoT (2).

To sum up when we will be embedding tokens to secure signaling messages and(or) voice transmission process will be as in a Fig. 5.

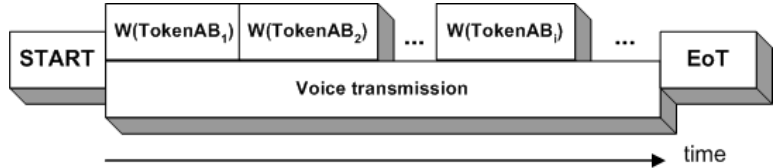

Figure 5 General transmission process

## 4  Conclusion and future work

In this paper new security architecture based on digital watermarking is proposed. It is totally new approach for securing signaling protocol's messages that are ex-changed between calling parties. We named it a *post factum* method because it works some time after the phase of exchanging of signaling messages took place. Nevertheless we find it very useful and flexible because this scheme is signaling protocol independent and potentially gives new possibilities for securing IP Telephony.

This solution has most advantages when we combine securing of signaling protocol and original voice. In this situation we can use one method to provide authentication and integrity for conversation and signaling messages simultaneously. This is especially important when in IP Telephony network multi-signaling protocols are used.

Future activities will be connected with comparison of latency and complexity for IP Telephony systems when using classical security architecture and our watermark based one. For example we can compare security mechanisms designated for SIP: SIP Digest or for H.323: Procedure I with our solution. Both mechanisms (SIP Digest and Procedure I) provide only authentication and integrity security services (without confidentiality) so they are the same security services as provided by our method.